# Metabolic Allometric Scaling of Unicellular Organisms as a Product of Selection Guided by Optimization of Nutrients Distribution in Food Chains

Yuri K. Shestopaloff

*Consultant, Prof. Dr. Sci. (Phys.)*
*Toronto, Ontario, Canada*
*shes169@yahoo.ca*

One of the major characteristics of living organisms is metabolic rate — the amount of energy produced per unit of time. When the mass of organisms increases, the metabolic rate also increases (as a power function of mass), but usually slower than mass. This effect is called metabolic allometric scaling. Its causes are considered unknown. The effect has important implications for individual and population organismal development. It was shown in the first part of this study, presented in a separate paper, that in the case of *multicellular* organisms, this effect is a consequence of natural selection and optimization of nutrient distribution between the species of a food chain, sharing resources of a common habitat. Here, in the second part that studies *unicellular* organisms, we discover that the same principle of natural selection guided by optimization of nutrient distribution between the species of a food chain defines also metabolic allometric scaling of unicellular organisms. To find that, we consider the metabolic properties of *Amoeba proteus*, fission yeast *Schizosaccharomyces pombe*, *Escherichia coli*, *Bacillus subtilis*, *Staphylococcus*. The sharing of nutrients is optimized in such a way that bigger microorganisms have progressively bigger nutrient influx per unit of surface. This evolutionary arrangement secures the stability of a food chain by providing certain metabolic advantages for bigger organisms. Accounting for this regular increase of nutrient influx with mass increase, we obtained allometric exponents and their ranges close to experimental values, thus proving that metabolic allometric scaling of both multicellular and unicellular organisms is defined by the same fundamental evolutionary principle of optimized sharing of nutrients between the species of a food chain.

## 1. Introduction

In order to support their life cycle, living organisms in an active state have to produce energy. The rate of energy production is called the metabolic rate. Metabolism of living organisms can be defined by emphasizing different aspects of this complex phenomenon. For our purposes, the definition in Webster's Dictionary is an adequate one: "The chemical changes in living cells by which energy is provided for vital processes and activities and new material is assimilated to repair the waste."

Metabolic rate generally increases slower than the mass of organisms.[1–5] (In some instances although the reverse relationships were observed, like in the case of bacterioplankton.[6]) This phenomenon is called metabolic allometric scaling. When it is considered across different taxa, this is *interspecific* allometric scaling (which is studied in this paper). When the same species are considered, the term *intraspecific* allometric scaling is used. The fundamental causes of interspecific allometric scaling are considered unknown. There are some credible hypotheses regarding the causes of intraspecific allometric scaling. According to Ref. 7, it relates to cellular properties influenced by heat dissipation specifics of organisms. This result is probably correct since other similar studies also point in the direction of cellular properties as key factors defining intraspecific allometric scaling. For instance, in Ref. 6, the authors acknowledge: "We have shown that cell size is a key functional trait in bacterioplankton communities." Despite the outer similarity of these two types of allometric scaling, the underlying mechanisms, defining these two phenomena, are rather different.[7]

This paper studies the metabolic allometric scaling of *unicellular* organisms. Allometric scaling of *multicellular* organisms was considered in a separate article,[8] which should be read first. Both articles *independently* discover the same general principle in their respective domains that the origin of metabolic allometric scaling is a product of natural selection and optimization of nutrient distribution between the species, composing a food chain within the same habitat.

## 2. Mathematical Description of a Metabolic Allometric Scaling Effect

Quantitatively, metabolism is characterized by the amount of produced energy per unit of time, called a metabolic rate ($B$). From the physical perspective, the metabolic rate is *power*, measured in *watts*. The dependence of metabolic rate from mass $M$ is often well approximated by a power function as follows:

$$B = aM^b, \tag{1}$$

where $a$ is a constant; the value of $b$ is called an *allometric exponent*; in most instances, $b < 1$.

Similar power function (1) is applicable to many other features of living organisms.[1,9–12]

Applying the logarithmic function to both the left and right parts of Eq. (1), one obtains the following equation:

$$\ln B = \ln a + b \ln M.$$

In other words, in logarithmic coordinates, ln $B$ is a *linear* function of ln $M$. Of course, this is only a mathematical approximation of the real phenomenon, and there are situations, when the value of $b$ varies with mass too, and we will consider such scenarios as well. Often, nonetheless, Eq. (1) gives a fairly accurate description.

The metabolic allometric effect was discovered in the 19th century by Rubner. In 1932, Kleiber stated two important properties of interspecific allometric scaling. The first was the formulation of Eq. (1). The second was estimation of allometric exponent $b$ (for animals within a wide range of masses) at the value of about 0.74, rounded for convenience to 3/4.

The history of this (in certain aspects fascinating) problem by itself is an interesting subject. A detailed excurse can be found in many works, including Refs. 1 and 9–12.

## 3. Conceptual Approaches to the Problem of Interspecific Allometric Scaling

Historically, the problem of interspecific allometric scaling was approached from several directions. Broadly, there were two main groups of hypotheses. One group includes hypotheses primarily asserting that certain physiological and physical constraints universally define the value of metabolic allometric scaling across different taxa. Examples include needs for heat dissipation, nutrient influx and waste removal limitations, imposed by surface–area relationships,[9–12] certain structures of distribution networks.[13,14]

The second group of hypotheses stems from the idea that it is natural selection, which acts through multiple adaptation and optimization mechanisms of physiological, physical, environmental nature and their interaction, which is responsible for the observed regularities of metabolic allometric scaling.

The results of our study of metabolic allometric scaling for unicellular organisms were first presented in the 2016-year version of this paper.[15] Results of similar study for multicellular organisms, which discovered the same foundational principle guiding evolutionary development of an entire food chain, were also reported in 2016. (The updated version of the last study was recently published as a separate article, see Ref. 8). Both studies were based on the idea that metabolic allometric scaling is created by natural selection within the concurrent boundaries, imposed by physical, environmental, physiological, organismal constraints (which, in turn, can change in time and space themselves). A distinguishing important aspect of the author's studies, however, was that unlike in other works, such a selection process occurs within a *food chain.* Species, composing such a food chain, occupy a common habitat and share their resources in a way, that is *optimal* for the entire food chain in that regard that it allows continuous reproduction of species within it in quantities, sufficient for a dynamic preservation of this food chain, while not allowing to acquire some species too much resources, which could jeopardize existence of other species of the food chain. It is this optimized by natural selection sharing of resources, which establishes the quantitative characteristics of metabolic allometric scaling, by securing sufficient resources for each species in the food chain to reproduce in quantities preserving the continuity of a food chain. On the other hand, the energetic advantage (expressed in different forms, like velocity, mechanical power) of one species over the others in the food chain cannot be too strong to lead to the extinction of species the bigger ones prey on or share common resources with. Otherwise, the food chain will be destroyed, and a new one should be created. This delicate, dynamic in nature, balance of resource sharing is what establishes equilibrium values of metabolic allometric scaling, observed in experimental studies, in a certain regular way.

However, at that time, eight years ago, the author's ideas were outside mainstream paradigms. The situation is changing now, and more researchers look at natural selection as a more plausible cause of metabolic allometric scaling, than particular physiological constraints. It would be premature to say that there is (or soon will be) a consensus even among researchers, considering natural selection as a primary cause of metabolic allometric scaling, about the main factors playing a role in such a selection and optimization process. However, the important thing is that all of them move away from the once popular idea about the definitive role of particular physiological mechanisms. (As a side note, recognition of really important new paradigms and discoveries might take decades. In this regard, the author may contrast not many historical examples, one of which was an immediate recognition by D. Hilbert of the famous Gödel's theorems. But that was Hilbert.)

The community of these researchers presently agrees on the following:

(a) Allometric scaling is considered as an integral effect produced by cooperative actions of multiple causes and mechanisms (physiological, physical, environmental ones), working within the common context created by natural selection and optimization. Examples of such works are Refs. 4, 5, 16 and 22.

(b) Different taxa may have different allometric exponents, arising from workings of natural selection and optimization on different combinations of factors and mechanisms defining this phenomenon.[3–5,9,17–22]

It is worth emphasizing that each such work arrives at a conclusion about the important role of natural selection in allometric scaling on the basis of different considerations and approaches, which makes the same final result more trustworthy. The main underlying ideas nonetheless are more or less universal: Living organisms can "overwrite" constraints imposed by particular mechanisms, and the leading factors in organisms' development are the environment, whose characteristics living organisms must fit in order to reproduce, and the availability of resources and the way living organisms handle them. Detailed analysis of such approaches is presented in the paper about allometric scaling in multicellular organisms.[8]

## 4. Metabolic Properties of Unicellular Organisms

### 4.1. *Metabolic allometric scaling as a product of natural selection and optimization*

The overwhelming majority of works about metabolic allometric scaling studied multicellular organisms. Metabolic properties of unicellular organisms have some specifics, which we discuss in this section, but also certain commonalities with the allometric scaling of multicellular organisms. First of all, this is the same action of natural selection and optimization, which define metabolic allometric scaling in multicellular organisms. We can find clear indications of this across many publications. For example, in Ref. 23, the authors acknowledge: "We show that allometric scaling relationships at lower levels of biological organization, such as body-size scaling of nutrient uptake and predation, give rise to scaling relationships at the food web and ecosystem levels." They continue: "Our findings explicitly connect scaling relationships at different levels of biological organization to ecological and evolutionary mechanisms." The aforementioned transition of scaling relationships from lower levels of biological organization to the food web and ecosystem levels is in line with our idea that metabolic allometric scaling is a product of optimized resource distribution across the food chain.

If the assumption about the primary role of natural selection in defining metabolic allometric scaling is true, then, given the great variety of factors, influencing different mechanisms in unicellular organisms, and environments in which unicellular organisms exist, the values of metabolic allometric scaling should vary substantially accordingly. Indeed, this is what many publications confirm. In Ref. 6, we find the following: "Our model also concludes that this general scaling exponent is not universal, with significant deviations along the transect."

Li and Wang[24] demonstrate the variability of allometric exponents through the dependence of both microbial metabolism and growth from the availability of resources: "Thus, we conclude that natural logarithmic microbial metabolism ($\ln(\lambda)$) and growth (or biomass) ($\ln(M)$) are both dependent on limiting resources."

Hatton *et al.*[25] go even further, explicitly stating the subjugated role of metabolism as a *server* of growth needs. The authors acknowledge: "Our findings are incompatible with a metabolic basis for growth scaling and instead point to growth dynamics as foundational to biological scaling. . . .These scaling laws encompass such core ecological characteristics as metabolism, abundance, growth and mortality." They continue: "We propose that *rather than limiting growth*, *metabolism adjusts to the needs of growth* (italics by YS) within major groups, and that growth dynamics may offer a viable theoretical basis to biological scaling."

Glazier[9] provides a detailed overview of the substantial variability of allometric exponents depending on the cell size and mode of growth (cell multiplication — including growth with cell size reduction, cell expansion and a combination of both). The conclusion the author draws is that such variability is a product of "Darwinian" natural selection within the context of multiple internal and external factors and mechanisms. A similar statement was presented in Ref. 11 by the same author.

Together, the quoted works, although each in its own way, point in the same direction that the metabolism of unicellular organisms is defined by natural selection and optimization in the space of different factors and their combinations, so that the entire process serves the reproduction needs of organisms by adjusting metabolism accordingly, but not the other way around.

### 4.2. *Effect of greater allometric exponents at the beginning of growth and in smaller cells*

García *et al.*[6] present results of its own research of oceanic plankton, and Glazier[9] reviews the results of different studies, which unambiguously demonstrate that allometric exponents generally are noticeably greater at the beginning of growth, and in smaller cells, while toward the end of growth their values reduce. This is not a universal phenomenon although, and there are exceptions. Based on the aforementioned studies, such exceptions very likely relate to a cell size during growth. Organisms, whose growth is supported by cell multiplication through the entire growth cycle, are more likely to have about the same high value of allometric exponent till the end of growth. The author of the present work did a similar research for intraspecific allometric scaling, and likewise discovered that organisms growing by mostly cell multiplication have greater allometric exponents,[7] approaching isometric (close to one) values.

Given a noticeable body of studies, which encountered this (still considered unexplained) effect, we will provide such an explanation and show, how it naturally follows from the growth dependencies described by the general growth mechanism, which we use here for finding metabolic properties of unicellular organisms.

### 4.3. *Specifics of metabolic properties of unicellular organisms*

Below, we review some specific aspects of the metabolism of unicellular organisms on the basis of Ref. 26. In this work, the authors computed metabolic rates for prokaryotes, protists and metazoans, and provided hypotheses for explanation of the obtained results. They assume that the high value of the allometric exponent in prokaryotes (1.72 for inactive and 1.96 for active species) is due to the evolutionary increase of genome size, which allowed for a greater variety of synthesized proteins and greater metabolic rate. When this reserve was exhausted with the size increase (apparently, the authors suggest, due to limitations on nutrient intake through the surface), the evolutionary development went in the direction of increasing the number of energy-producing organelles throughout the entire volume of unicellular organisms, which provided about isometric allometric scaling. Then, the appearance of nutrient distribution networks in multicellular organisms, and associated transportation costs, take tall on energy resources, which led to the following reduction of allometric exponent to about 0.76.

Note that the overwhelming majority of studied prokaryotes in the graph from Ref. 26 occupy about *two* orders of magnitude from *sixteen*, forming sort of a well-localized cloud dataset. Data for the other two groups have large dispersion. In such a situation, the results of regression analysis could easily vary tens of percent and even several times, depending on the chosen dataset. If one takes data for larger prokaryotes and the rest of the data, then the regression for the whole dataset would produce the allometric exponent noticeably less than one. (In fact, such large dispersion of data for unicellular organisms is presented in other works, and this is rather a typical feature of these organisms, which very likely relates not to the accuracy of measurements, but to real dispersion of metabolic characteristics of unicellular organisms.)

The hypothesis about the increase of metabolic power in prokaryotes with the increase in the size of the genome looks feasible (if considered together with other factors; not as a single "all defining" mechanism). On the other hand, explanations of metabolic mechanisms for protists and metazoans are not convincing, given the large dispersion of presented data and strong dependence of numerical results on the chosen datasets. Besides, the authors still resort to a *single* mechanism, while the present consensus is that allometric scaling is rather a *multifactor* phenomenon. Indeed, the earlier mentioned studies show that different classes of animals have noticeably different allometric exponents. So, we should not disregard such a possibility for unicellular organisms too.

Thus, the study of allometric properties of unicellular organisms is far from complete. In this paper, we provide proof that the evolutionary development of microorganisms within the food chain is an important factor affecting allometric scaling. We show that two factors play an important role. The first one is the common nutritional environment, which imposes limitations on the possible nutrient intake. For unicellular organisms, this factor is of special importance, because unicellular organisms acquire nutrients directly from the environment through the surface (save for some variations such as endocytosis, which also involves the surface, but in a somewhat different way).

The second factor is a regular increase in the amount of nutrients per unit surface, when the size of organisms increases, as discovered in this study. We discuss how this factor relates to evolutionary development and life organization, since, as we found, such an evolutionary increase in size strongly correlates with the increase of energetic capabilities of an entire organism. One of the reasons for this phenomenon is an evolutionary requirement to successfully compete for nutrient acquisition needed for sustainable reproduction. We discuss why such an increase of acquired nutrients per unit surface in unicellular organisms is of so regular and persistent nature through the entire food chain.

This evolutionary energetic increase, apparently, exposes one of the important properties of evolution in general. By and large, all organisms are both prey and predators. Organisms adapt to the environment by different means, but an increase in size is one of the main evolutionary paths, around which other developments evolve. (Imagine a kind of evolution without the size increase. As Dr. K. Y. Shestopaloff noted with regard to evolution in general (not only to size), "evolution goes from bottom to top".) In a situation of a relatively stable food supply, it should be expected that the size increase brings certain advantages, and first of all energetic ones, in order to get food more reliably for successful reproduction. (It is an established fact that in the conditions we have had so far on the Earth during different geological periods, the increase of organismal size and related energetic capabilities was one of the major evolutionary paths. According to findings presented in this paper, this evolutionary trend, besides other effects, plays an important role in interspecific allometric scaling.)

So, we argue that this is rather the entire evolutionary process of organic life development within the food chain (supported by numerous physical, and physiological mechanisms and environmental conditions), which is largely responsible for the interspecific allometric scaling and its stability. Saying this, we do not mean some group selection (while neither deny such a possibility, for which this notion should be well defined), but the *Darwinian* natural selection of individual species under the influence of the same or similar environmental factors. Or, as an authoritative scientist in Ref. 27 put it, "... as statistical summations of the effects of individual adaptations". So, the proposition of this paper that species living in the same habitat and continuously interacting with each other (say, in a predator–prey relationship), influence the development of each other, should not cause objection. By and large, this is the only thing we are saying with regard to coevolving development of organisms within the same food chain.

## 5. Finding Metabolic Properties of Cells

### 5.1. *General growth mechanism and its application to finding metabolic properties of unicellular organisms*

In Refs. 28–30, a method for finding growth and metabolic characteristics of microorganisms was introduced and verified by experimental data for *Schizosaccharomyces pombe*, *Amoeba*, *Schizosaccharomyces cerevisiae*, and for few other unicellular organisms. The approach is based on the earlier discovered and presently considered a proven general growth mechanism, acting at higher than molecular levels. Its core property is that it defines the amount of nutrients used for biomass production, represented as a fraction of the total consumed nutrient influx. Thus, at any given moment of growth, we know how much nutrients are used for biomass synthesis and how much for organism's maintenance needs. This way, we also know the total amount of consumed nutrients, which correlates very well with the organismal metabolic rate.

When we say "nutrients", we mean all substances acquired by cells. The correlation of nutrient intake in unicellular organisms with their metabolic rates was discussed in Refs. 28 and 29 and in some other works on the subject by the same author. One of the reasons for such a high correlation is that cells have a single biochemical machinery, in which there are no separate biochemical machineries for biomass synthesis and maintenance, but all biochemical reactions interrelate so that the overwhelming amount of acquired nutrients are included in on-going chemical reactions.

Using the results of modeling the growth of unicellular organisms and found metabolic properties from Refs. 28–30, below we consider metabolic characteristics of unicellular organisms. Experimental data were taken from the following literature:

*S. pombe* — from Refs. 31 and 32;
*Bacillus subtilis* — from Ref. 33;
*Escherichia coli* — from Ref. 34;
*Amoeba proteus* — from Ref. 35.

The size (mass and volume) of considered organisms covers seven orders of volume magnitude, from $0.32\,\mu m^3$ for *B. subtilis* to $1.88 \cdot 10^7\ \mu m^3$ for the grown *Amoeba*. Cells grew in normal conditions, that is they were not exposed to starvation or other extreme regimes. We use a growth model from Refs. 29 and 30, adjusted to the geometry of considered organisms (details are in Ref. 30), thus finding nutrient influx for (a) the whole organisms; (b) per unit of volume; (c) per unit of surface. For *Staphylococcus*, we do not have experimental data, so we use the growth model alone. (For *Staphylococcus*, based on the results of modeling, we have made several predictions, which would be interesting to verify experimentally.) For validation of obtained results, we also use experimental results for excised cells growing *in vitro* in culture.[17,18]

The fact that we consider prokaryotes and eukaryotes in the same dataset, despite the differences between these organisms, has no effect on the results from the perspective of the general growth mechanism, which is applicable to unicellular organisms in general. Similarly, DeLong *et al.*[26] put different organisms into the same dataset just on the basis that they possess metabolic properties — the characteristic that was studied.

### 5.2. *Metabolic rates of S. pombe*

Figure 1 shows the nutrient influx (amount of nutrients per unit time) for *S. pombe* at the end of growth, obtained from experimental observations in Ref. 31. The metabolic rate per unit volume (the lower dataset) decreases with the increase of volume (for the constant density, we can substitute volume for mass). The regression line has the following parameters: intercept is equal to 1.569, the slope is equal to ($-0.99$), which means that the amount of nutrients per unit volume *decreases* as fast as the total volume *increases*. Thus, the assumption from Ref. 26 about the same energetic capability of units of volumes for protists is not confirmed in this case. Interestingly enough, this fact means that all considered fission yeast, regardless of their volume, obtain about the same total amount of nutrients per unit of time. Indeed, the regression line for the upper dataset, representing the nutrient influx for entire organisms, has a negligible slope of 0.0104 and the same intercept of 1.569.

Fig. 1. Nutrient influx in *S. pombe* versus volume, in logarithmic scale. Black diamonds denote metabolic rate per unit volume ($K_V$, measured in pg $\cdot$ $\mu$m$^{-3}$ $\cdot$ min$^{-1}$); diamonds denote the total metabolic rate ($K_T$, measured in pg $\cdot$ min$^{-1}$); "pg" stands for "picogram".

$p$-value is equal to 0.932, which means that all *S. pombe*, indeed, on average obtain the same amount of nutrients regardless of their size.

Experimental data for excised cells growing in cultures,[17,18] for mammalian hepatocytes, dermal fibroblasts, skeletal myoblasts and avian dermal fibroblasts show that except for the weak allometric scaling for hepatocytes, the rest of the cells show little allometric scaling depending on the body size in the range of masses of several orders of magnitude. We can be certain that the sizes of studied cells were different too, and so the above conclusion about weak allometric scaling is also valid with regard to the cell size. Indeed, such a characteristic behavior we observe in our data in Fig. 1 for the metabolic rate of whole unicellular organisms.

Therefore, the obtained result for *S. pombe* complies with experimental observations for other single cells, which adds credibility to our approach for finding the metabolic properties of single cells.

### 5.3. *Metabolic characteristics of unicellular organisms*

Similarly to *S. pombe*, we found metabolic characteristics for *E. coli* and *B. subtilis*, since they have the same rod-like shape. *Amoeba*'s nutrient consumption was found using the *Amoeba*'s cylinder-like growth model from Refs. 28 and 30, and experimental data from Ref. 35. *Staphylococcus*'s growth was modeled by a growing sphere. Results are summarized in Table 1 and Fig. 2. It's interesting to note that *all* data points for the total nutrient influx at the *end of growth* are located very close to a regression line (Fig. 2(a)). The regression line has a slope $b = 0.758 \pm 0.015$ and an intercept of $(-0.288)$.

Table 1. Change of nutrient influx during growth, and comparison between organisms. (NPUV stands for "nutrient influx per unit volume", MNIUS stands for "maximum nutrient influx per unit surface".)

| Organism | Volume increase, $\mu m^3$ | Total nutrient influx, $pg \cdot min^{-1}$ | NPUV at max. size $pg \cdot \mu m^{-3} \cdot min^{-1}$ | Relative increase NPUV during gr., times | NPUV less than in *E. coli* 1, times | MNIUS $pg \cdot \mu m^{-2} \cdot min^{-1}$ | MNIUS relative to *E. coli* 1, times | Addit. to allomet. expon., end of growth $b_n$ |
|---|---|---|---|---|---|---|---|---|
| *Amoeba* | (0.92–1.88)E+7 | 244,400 | 0.013 | 1.44 | **204** | 0.323–0.545 | **2.26** | |
| *S. pombe* | 166–325 | 63.4 | 0.195 | 6.9 | **13.6** | 0.03–0.222 | **0.92** | 0.082 |
| *E. coli* 1 | 1–4 | 10.6 | 2.65 | 12 | **1** | 0.021–0.284 | **1** | 0.0424 |
| *E. coli* 2 | 2.39–4.96 | 8.5 | 1.713 | 4.71 | **1.55** | 0.084–0.234 | **0.82** | 0.056 |
| *Staphylococcus* | 1.07–2.145 | 1.12 | 0.522 | 1.25 | **5.1** | 0.093–0.147 | **0.61** | 0.082 |
| *B. subtilis* | 0.32–0.617 | 0.66 | 1.066 | 3 | **2.5** | 0.043–0.134 | **0.56** | 0.0814 |
| Average | | | | | | | | **0.072** |

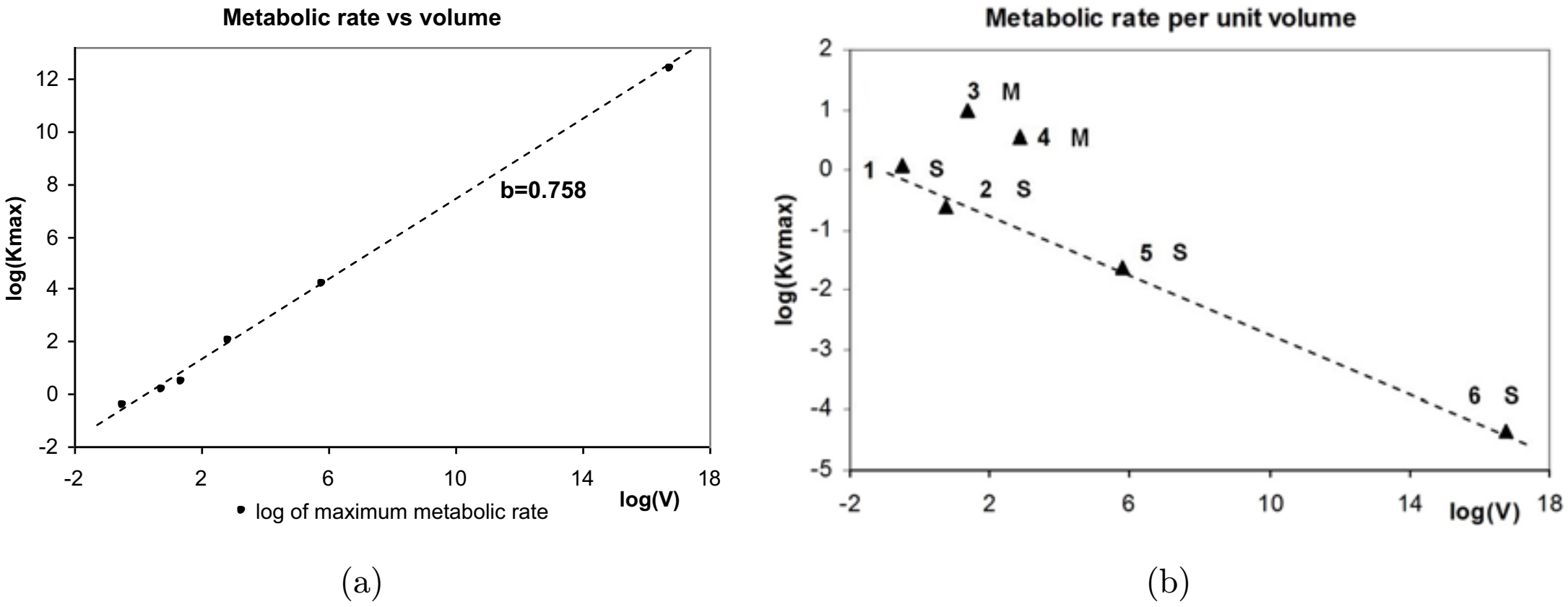

(a) (b)

Fig. 2. Change of metabolic rate depending on volume, in logarithmic scale. Numbered data points from left to right correspond to *B. subtilis* (1), *Staphylococcus* (2), *E. coli* 1 (3), *E. coli* 2 (4), *S. pombe* (5), *Amoeba* (6). *S* and *M* accordingly denote sedentary and motile microorganisms. Experimental data are from Ref. 33 for *B. subtilis*, from Ref. 34 for *E. coli*, from Ref. 31 for *S. pombe*, from Ref. 35 for *Amoeba*. (a) whole organisms; (b) per unit volume.

At the *beginning of growth*, the allometric exponent $b = 0.853 \pm 0.069$, that is significantly greater. This result, on the one hand, confirms the experimental effect earlier discovered in other studies and described in Sec. 4.2. On the other hand, our result explains the origin of this effect, which is a higher metabolic rate at the beginning of growth of *all* considered microorganisms. Thus, indeed, one has to distinguish the allometric scaling at the beginning and at the end of the growth cycle.

This result is also useful in that regard that it explicitly confirms in a *quantitative* form that the allometric exponents can significantly vary not only across different taxonomic groups, but also ontogenetically. Furthermore, this variation is not confined to a single microorganism, but is universally common, across at least five different microorganisms, which we considered. In turn, this variability further enforces the idea that the primary cause of allometric scaling is natural selection and optimization within the constraints, imposed by physical, physiological, and environmental factors.

## 5.4. *Variations of nutrient influx*

### 5.4.1. *Nutrient influx per unit volume*

Using the same growth models as above, we found that unicellular organisms are able to substantially increase nutrient influx during growth — faster than mass's increase. This theoretical finding is very well supported by experimental observation of such "superlinear scaling of microbial metabolic rates" described in Ref. 6 and other works. On one hand, this result provides one more proof of the validity of the growth models we used here, and on the other hand, explains the origin of this experimentally observed effect, which is a higher metabolic rate at the beginning of growth.

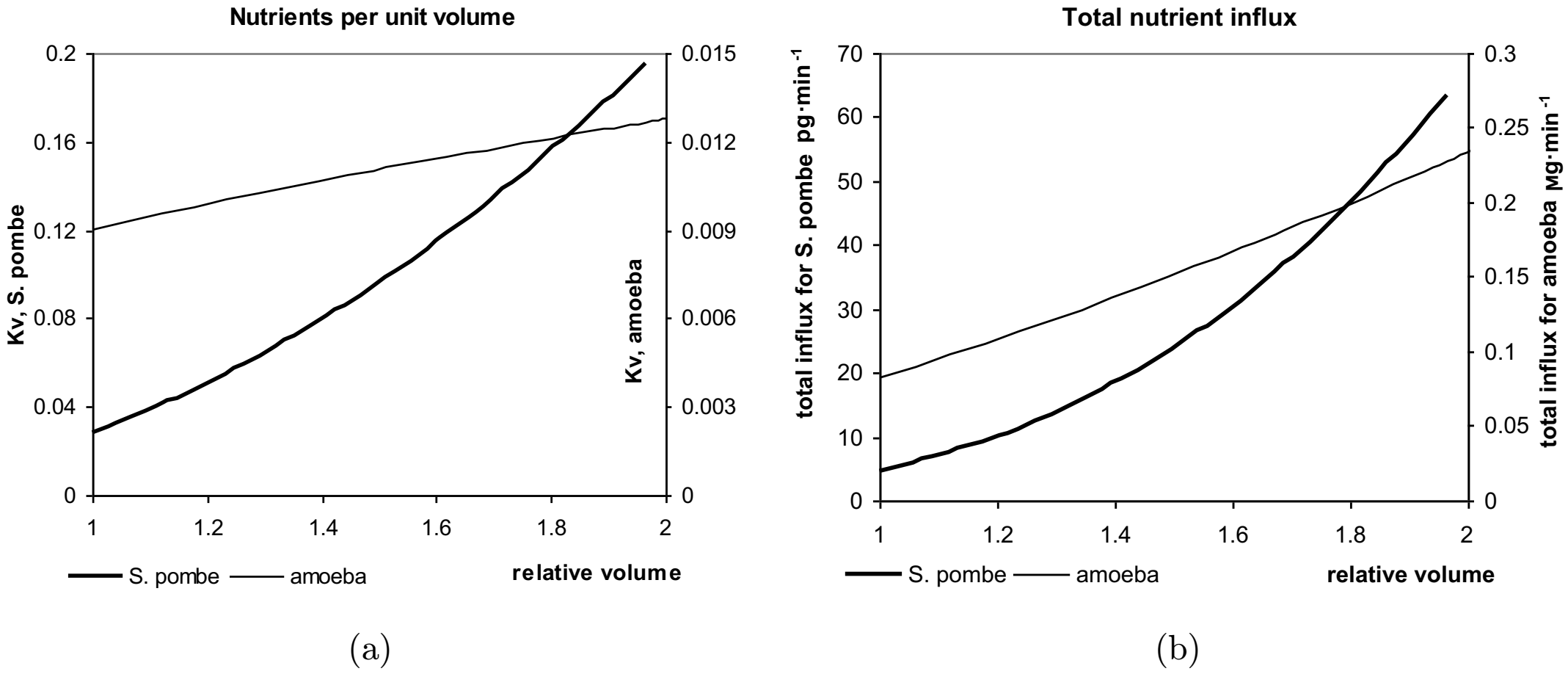

Fig. 3. Nutrient influx in *S. pombe* and *Amoeba*. (a) Per unit volume, units of measure are $\text{pg} \cdot \mu\text{m}^{-3} \cdot \text{min}^{-1}$. (b) Total nutrient influx.

Figure 3(a) and Table 1 present the increase of nutrient influx per unit volume during growth of a *single* organism. We can see that the increase for *S. pombe* is about 6.9 times, and for *Amoeba* it is 1.44 times. A similar graph for *E. coli 1* would produce a 12-fold increase.

However, an even more dramatic difference in nutrient influx per unit volume is observed *between* organisms (Table 1). *Amoeba* consumes 204 times less nutrients per unit volume than *E. coli* 1.

Figure 3(b) shows the change in nutrient influx for the entire organisms. Although in both instances volume increased by about two times (except for *E. coli* 1), the total nutrient influx increased by 13.5 times for *S. pombe* and 2.8 times for *Amoeba*. In the case of *E. coli* 1, this increase was 54.3 times, versus the four times volume increase.

As we can see from Table 1, nutrient influx per unit volume in different unicellular organisms can vary by as much as *hundreds times*, and ten times during ontogenetic development. In other words, unicellular organisms demonstrate very wide range of the total nutrient influx, influx per unit volume and across different taxa and ontogenetically.

5.4.2. *Specifics of nutrient influx through surface*

However, if we consider variations of nutrient influxes *per unit surface*, then we see that they are by two orders of magnitude less, as Table 1 shows (the second and the third columns from the right). *Amoeba*'s nutrient influx per unit surface is greater than in *E. coli* 1 only by 2.26 times, while the nutrient influx per unit volume is greater in *E. coli* 1 than in *Amoeba* by 204 times. For *S. pombe* and *E. coli*, nutrient influxes per unit surface differ by only 8%, despite the great difference, of 81 times, in the sizes of these organisms. (For *S. pombe*, we chose the experimental curve at 32°C from Ref. 31, which has the least dispersion of data. Other experimental dependencies produce similar results.)

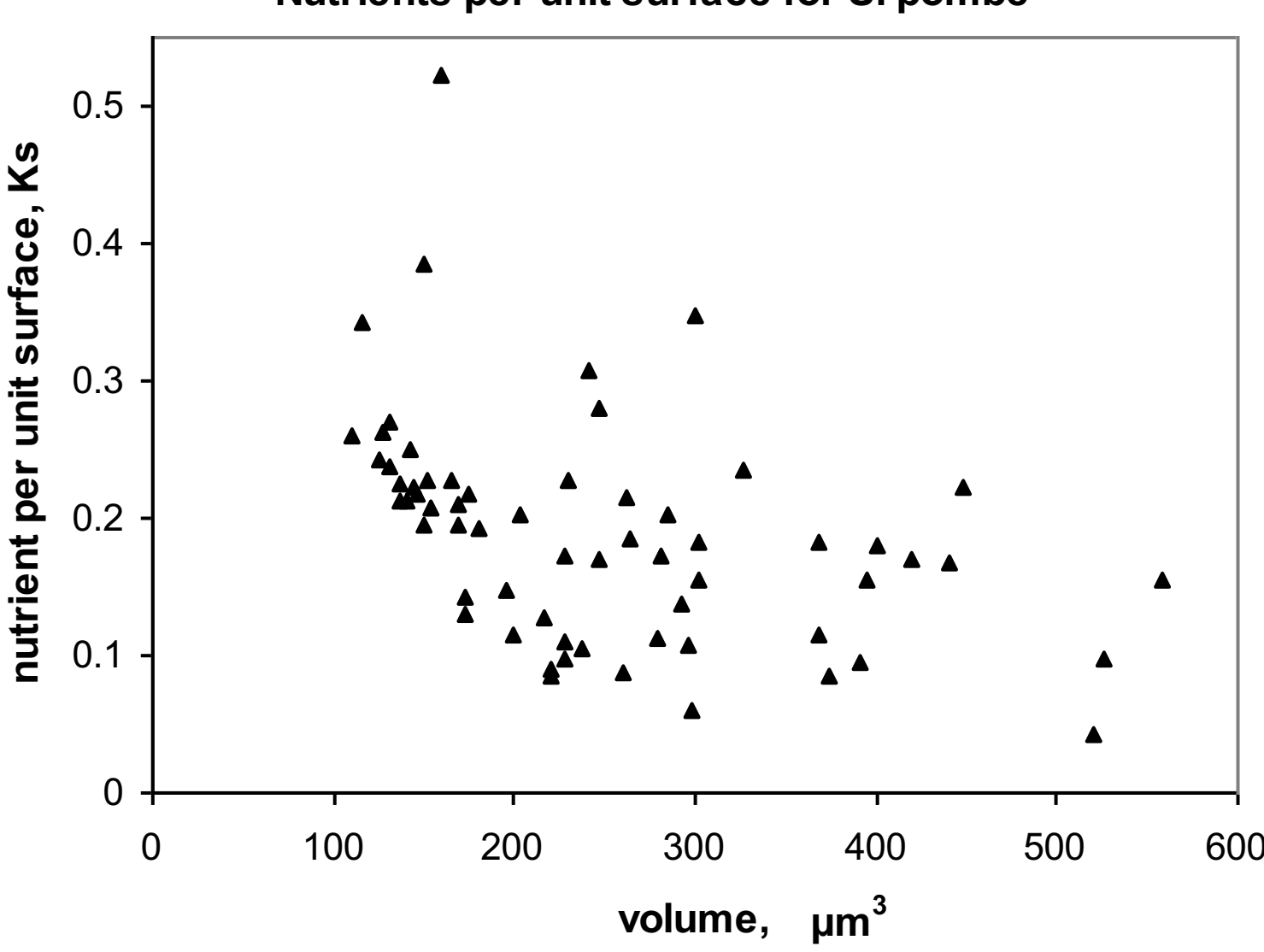

Fig. 4. Nutrient influx per unit surface for *S. pombe* depending on volume, for the data from Ref. 31. Units of measure are $\mathrm{pg} \cdot \mu\mathrm{m}^{-2} \cdot \mathrm{min}^{-1}$.

Note the variations of nutrient influx per unit surface for the same species. Figure 4 shows such variations for *S. pombe*. For two experimental observations of *E. coli* from Ref. 34, we also obtained different values of nutrient influx per unit surface (0.284 and $0.234\,\mathrm{pg} \cdot \mu\mathrm{m}^{-2} \cdot \mathrm{min}^{-1}$). All nutrient influxes per unit surface found for considered organisms are limited to the range of about $0.08$–$0.55\,\mathrm{pg} \cdot \mu\mathrm{m}^{-2} \cdot \mathrm{min}^{-1}$. As Fig. 4 shows, *S. pombe*, if we include marginal values, covers almost entirely this range, although the majority of data is in the range of $0.088$–$0.28\,\mathrm{pg} \cdot \mu\mathrm{m}^{-2} \cdot \mathrm{min}^{-1}$. This result suggests that other organisms may have similar ranges of nutrient consumptions per unit surface, depending on combination of organismal characteristics and environmental conditions.

Discovered variations of nutrient influxes per unit volume and per unit surface mean accordingly very wide range of adaptive abilities of considered unicellular organisms. Nonetheless, their metabolic rates show a well ordered allometric regularity (with an allometric exponent of about 0.758 at the end of growth and 0.853 at the beginning of growth). Thus, there should be influential factors, which force such a regular change of metabolic rates, and these factors should not be only intrinsic ones, since, as we have seen, the core metabolic characteristics vary widely.

## 6. Accounting for Changes in Nutrient Acquisition Between Developmental Stages

Let us consider evolutionary change in nutrient acquisition. We assume that there were $x$ hypothetical successive evolutionary development stages, each producing a bigger organism than the previous one. The relative mass increase is by $g$ times at each stage, so that

$$M_x = M_0 g^x. \tag{2}$$

Let the amount of consumed nutrients $F$ increasing by $g^b$ times per development phase. (We can use the power function because the existence of an allometric exponent is an empirical fact.) Then, the amount of nutrients $F_x$ consumed at $x$ stage, is as follows:

$$F_x = F_0 g^{bx}, \tag{3}$$

where $F_0$ is the amount of nutrients at the first stage.

Substituting the value of $g^x$ from Eq. (2) into Eq. (3), we obtain

$$F_x = F_0 (M_x/M_0)^b. \tag{4}$$

We assume that the density is constant, so that mass is proportional to volume $W$, $M \propto W$, and consequently $M_x/M_0 = W_x/W_0$. Note that both the number of developmental stages $x$ and the mass increase $g$ per stage disappeared in Eq. (4). In fact, we obtained an equation very similar to Eq. (1) for the metabolic rate.

### 6.1. *Factors defining allometric scaling in unicellular organisms and its variability*

The analysis above showed that there are at least two factors defining the allometric scaling of unicellular organisms. The first one is the consequence of nutrient acquisition through the surface, which depends on nutrients' availability, the surface of organisms and their metabolic needs. The last ones, as we found, can noticeably vary even for the same species. The mode of motion and other specific characteristics of organisms also contribute to significant variations of metabolic rate for organisms with similar sizes.

Nutrients acquired through the *surface* are processed by *volume*. For three-dimensional (3D) increasing organisms, it means the allometric exponent of 2/3 (surface is proportional to a square of a linear size, while volume is proportional to a cube of linear dimension, from which the value of 2/3 follows). We denote this allometric base exponent as $b_s = 2/3$ ("$s$" stands for "surface").

The second factor is the discovered regular increase of nutrient influx per unit surface with the growth of mass (Table 1). We will denote this allometric exponent as $b_n$ ("$n$" stands for "nutrients"). It can be found as follows. Let us denote the amount of nutrients per unit surface as $k_A$, indexes "1" and "2" correspond to two different organisms. Then, according to Eq. (4)

$$(k_{A1}/k_{A2}) = (M_1/M_2)^{b_n}. \tag{5}$$

The solution of Eq. (5) is as follows:

$$b_n = \ln(k_{A1}/k_{A2})/\ln(M_1/M_2). \tag{6}$$

The resulting allometric exponent $b$ then can be found as the sum of $b_s$ and $b_n$, which follows from the equation as follows (we use an equivalent form of presenting Eq. (1), discussed in Ref. 8):

$$B = B_2(M_1/M_2)^{b_s}(M_1/M_2)^{b_n} = B_0(M_1/M_2)^{b_s+b_n}. \tag{7}$$

We can use Eq. (6) in order to compute values of $b_n$ for each species of unicellular organisms relative to *Amoeba* (using the same data, which earlier produced the values of allometric exponent for the end of growth $b = 0.758 \pm 0.015$, and $b = 0.853 \pm 0.069$ for the beginning of growth). Results are shown in the last column of Table 1. The linear regression produces accordingly the value of $b_n = 0.0724 \pm 0.013$ with intercept of $(-0.174)$. Adding the base allometric exponent $b_s = 2/3$, corresponding to the surface–volume relationships for 3D increasing organisms, we obtain $b = 0.739 \pm 0.013$. This number is close to the value of $0.758 \pm 0.015$ found on the basis of experimental observations, for the end of growth. The slightly lower value of the obtained allometric exponent is mostly due to the high energetic abilities of *E. coli*, which uses the energy-demanding mode of locomotion — swimming.

The result we have just obtained is a remarkable one indeed. Here, we have made an assumption that the allometric exponent of microorganisms is due to the balance achieved through natural selection and optimization in sharing common resources, nutrients, in the nutritional environment common for all considered unicellular organisms. Then, we compared the thus obtained result with a value of allometric exponent deduced from experimental data, and found that these two values are indeed very close! What better direct proof can be of the primary role of natural selection in establishing values of allometric scaling by finding an optimal distribution of shared nutritional resources across the food chain of unicellular organisms?

All other organisms except *E. coli* have *low motility*, which explains the difference in the values of nutrient influx per unit surface between them and *E. coli*. For these sedentary organisms, Eq. (6) produces very close values of $b_n$. The regression line has a slope of $b_n = 0.0814 \pm 0.0006$, which corresponds to the allometric exponent $b = 0.748$. Figure 5 shows this graphically. Such consistency says in favor of the validity of the proposed explanation of the allometric scaling phenomena in the studied unicellular organisms, according to which the resulting allometric exponent is composed of the base "surface–volume" exponent of $2/3$, plus the addition due to a regular increase of nutrient influx per unit surface, when the size of organisms increases. (Although, as we could see, such an increase also depends on the locomotion mode — actively moving organisms, such as *E. coli*, need more energy and accordingly more nutrients.) If we think for a moment, that should not come as a surprise at all, since the only two factors upon which the amount of acquired nutrients in unicellular organisms depends are the rate of nutrient acquisition through the unit of surface, and the surface area.

As we found, the actual nutrient acquisition per unit surface can noticeably deviate from average values (see Table 1 and Fig. 2). *E. coli* 1 consumes more nutrients per unit surface than the chosen *S. pombe* (although on average *S. pombe* consume substantially more than *E. coli* 1, as it follows from Fig. 4, so that the notion about the regular increase of average nutrient intake per unit surface with the increase of organisms' mass remains valid). If we compute the allometric exponent of *E. coli* 1 relative to *S. pombe*, then we obtain the value of 0.611 (which is, interestingly enough, close to the lower limit of the range 0.608–1.09, reported in Ref. 10). At the beginning of growth, new cells at normal growth conditions have substantially higher values of allometric exponents (the average value in our case was $b = 0.853 \pm 0.069$), which is due to intensive biomass synthesis at this phase of growth.

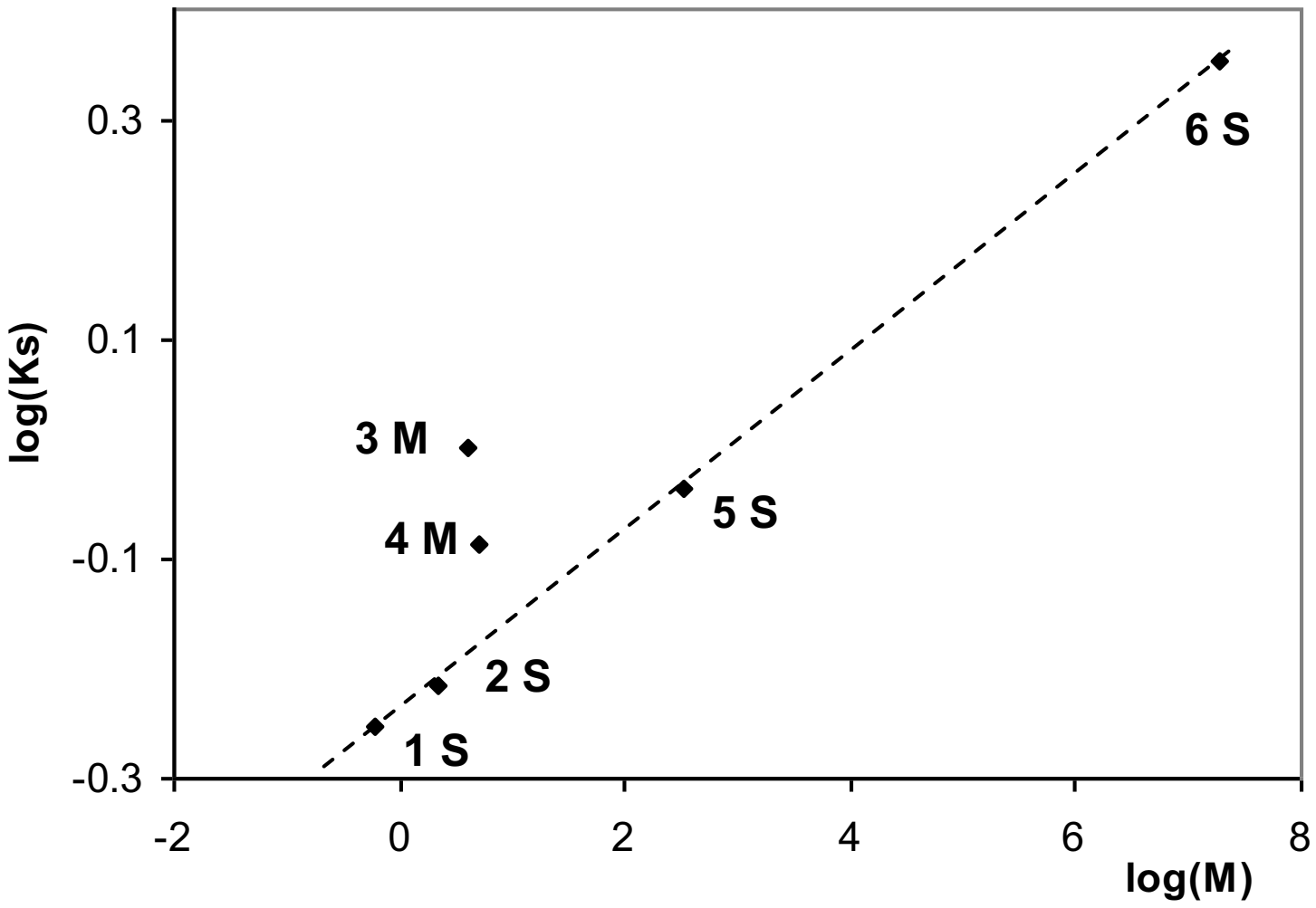

Fig. 5. Nutrient influx per unit surface depending on organisms' mass, in logarithmic scale. Numbered data points correspond to *B. subtilis* (1), *Staphylococcus* (2), *E. coli* 1 (3), *E. coli* 2 (4), *S. pombe* (5), *Amoeba* (6). $S$ and $M$ denote accordingly sedentary and motile organisms. Nutrient influx per unit surface $K_s$ is measured in pg $\cdot\, \mu\text{m}^{-2} \cdot \text{min}^{-1}$, mass $M$ is measured in picograms (pg).

So, the range of variability, which we obtained for our data using the proposed approach, is commensurate with the known variability of allometric exponents for unicellular organisms. The range depends on the physiology of organisms, phase of their life cycle, mode of motion, specifics of biochemical mechanisms, as well as nutrients' availability and composition, temperature and other environmental factors. Depending on nutrients' availability, the same organisms can use aerobic or anaerobic metabolic pathways, and accordingly may have different metabolic characteristics. In our study, we found that allometric exponents, for all considered organisms, substantially differ depending on the phase of growth, with high values of about 0.853 at the beginning of growth versus 0.758 at the end of the growth cycle. Results, obtained in Ref. 31, also show a strong dependence of metabolic rate in fission yeast on temperature. So, our approach by no means contradicts available results, but rather unites them under one umbrella on the basis of two principal factors, whose variations provide the range of metabolic properties corresponding to available experimental data.

Table 2. Calculated allometric exponents and their ranges versus experimental data.

| | End of growth, calculated | End of growth, experiment | Beginning of growth, calculated | Beginning of growth, experiment |
|---|---|---|---|---|
| Value of allom. exponent | $0.739 \pm 0.013$ | $0.758 \pm 0.015$ | $0.813 \pm 0.061$ | $0.853 \pm 0.069$ |
| Range of allom. exponent | 0.611–0.759 | 0.743–0.773 | 0.752–0.874 | 0.784–0.922 |

For the beginning of growth, using Eq. (6), we obtained $b = 0.813 \pm 0.061$, versus the value of $b = 0.853 \pm 0.069$ found on the basis of experiments. We already mentioned that the beginning of growth is characterized by very different start conditions, so the standard deviation should be higher than at the end of growth, and it is. We can see that both found ranges significantly overlap.

As Fig. 4 shows, nutrient influx per unit surface can significantly vary for the same species, and also during the growth cycle. Unfortunately, we have statistically meaningful data for *S. pombe* only. However, even considering these data, we found that the variations of the allometric exponent for *S. pombe* relative to *Amoeba* are in the range 0.7–0.91 for the end of growth (experimental points corresponding to volumes 150 and 521 $\mu m^3$ and nutrient influxes of 0.384 and 0.043 $\text{pg} \cdot \mu\text{m}^{-2} \cdot \text{min}^{-1}$). Similar variations of the allometric exponent for *B. subtilis* relative to *S. pombe* are in the range of 0.66–0.86. In other words, only variations of nutrient influx for the single species can provide the range of allometric exponents commensurate with the results derived from experimental observations.

Thus, our explanation of the mechanism of allometric scaling in unicellular organisms is consistent for microorganisms with regard to the known range of allometric exponents too. Table 2 summarizes the main results. Overall, we obtained the range of allometric exponents for the studied organisms of 0.611–0.922, which is in good agreement with the range of 0.608–1.09 from Ref. 10. We should emphasize one more time that available data for unicellular organisms are characterized by wide variations, so that such a wide range should not be considered as large errors, but rather as specifics of metabolic properties of unicellular organisms.

The main idea with regard to variability is this. As in every natural phenomenon, there are main mechanisms, which define the core properties of the phenomenon, while other factors of lesser prominence *modulate* these properties within certain ranges. This is what we eventually accomplished — we discovered the principal mechanisms of allometric scaling in unicellular organisms and showed that the influence of secondary factors, namely the ordered with mass increase variations in nutrient influx per unit surface, provide the range of allometric exponents commensurate with available experimental observations.

## 7. Discussion

In the other author's work,[8] studying the origin of the metabolic allometric phenomenon in multicellular organisms, and considering mammals, reptiles, birds and fishes, it was found that the metabolic allometric scaling is also due to several simultaneously acting factors, and the same evolutionary adaptation within the food chain, when each next increase in size leads to increase of metabolic rate above the base allometric exponent, defined by biomechanical constraints. The described mechanism in the case of multicellular organisms' mirrors one-to-one what we discovered for unicellular organisms: There is a base allometric exponent, and the addition is due to the increase of nutrient influx per unit surface when the size of microorganisms increases.

Thus, the same principle, which is a consequence of an evolutionarily established dynamic balance of a food chain, turned out to be the major determinant, both in the realms of unicellular and multicellular organisms.

In the case of mammals, the links in the food chain are more visible, since the predator directly feeds on the neighboring preys, while unicellular organisms relate to each other both through the common nutritional environment, and also by feeding on each other. Nonetheless, their "predator–prey" relationships are not so obvious, because of the diverse forms of existence, ways of feeding and modes of motion.

Nature is a virtuoso, capable of creating forms of life, acquiring nutrients sometimes in very exotic ways, to which a great diversity of environmental conditions contributed. Organisms use all possible means to acquire nutrients for successful reproduction, whatever is available in their disposition biochemically, physiologically and from the environment. This is always an interplay of many factors, which defines, who eats who. However, behind these masking the real development scenarios, there is an evolutionary backbone, which is the increase of organisms' mass and transformations associated with it, including metabolic rates. This is not a straightforward development, but a complex interplay depending on many factors, with possible rollbacks. However, at the core, it is *this* evolutionary backbone, from which other secondary evolutionary branches start (which, in turn, sometimes produce bewildering mechanisms and means of nutrient acquisition).

Very similar (and similarly important) considerations are applicable to unicellular organisms. All unicellular organisms compete for nutrients. Bigger organisms originate in the same nutritional environment because they are more successful in nutrient acquisition. In other words, they already have greater metabolic capacities. Gradually, such organisms can evolve into different species. However, what is important to remember, the basis of their larger size is the ability to acquire more nutrients from the environment compared to competitors; without that, they would not survive as bigger organisms. Since unicellular organisms acquire nutrients through the surface, then such an advantage should be expressed as the ability to get more nutrients through the surface (and namely per unit of surface), from the same environment where competitors reside. If we think for a moment, that is the most logical way: on one hand, this means acquiring metabolic advantage; on the other hand, this means getting nutrients quicker than competitors in the same environment, thus depriving them of food to the extent, which begins affecting the size of their populations. Such a "population management" effect can be achieved most reliably if the nutrient acquisition in bigger species *per unit surface* is greater;

otherwise, the accumulated amount of nutrients, acquired from the environment by smaller organisms, will increase faster than that by the bigger organisms, putting the last ones in nutritional disadvantage. Since nutrient acquisition per unit surface can substantially vary (which we saw in case of *S. pombe*, Fig. 4), it means that these advantageous relationships can be very sensibly and accurately regulated and adjusted to keep the overall balance of the food chain through numerous feedback loops. This argument is further enforced by the consideration that such a broad adjusting capability is rather not an option but a necessity for unicellular organisms, experiencing a wide range of environmental conditions, like changes in temperature, nutrient concentration, etc.

Nature shows examples of amazingly energetic creatures, both aquatic and terrestrial. Many species show isometric intraspecific allometric scaling; for instance, ants (Ref. 36), many fishes, squids (Refs. 10 and 11). Mammals show high adaptability to metabolic requirements. Depending on the situation, they may also show isometric scaling. Reviewing the previous works,[37] the authors indicate that "the cost of climbing 1 m was nearly the same per kilogram, regardless of weight", which means isometric allometric scaling, while the same animals show significantly smaller allometric exponents in overtaking horizontal distances. Our results above also show substantially higher metabolic capacity of *E. coli* compared to other considered organisms in their size range. So, the biochemical limitations on the upper level of metabolic activity, in principle, are not so restrictive as we observe in case of interspecific allometric scaling.

This is something else that keeps the increase of metabolism in progressively bigger species checked, both in unicellular and multicellular organisms. For multicellular organisms, we found proofs that this upper limit is imposed by the need to keep the *integrity* and *continuity* of an entire food chain. Shestopaloff[38] demonstrates numerical examples that populations of species are very sensitive to food availability, and that even insignificant fluctuations in food supply can cause drastic fluctuations in population quantity, including extinction scenarios. Thus, evolutionarily, the entire food chain is kept in a dynamic balance, filtering both too aggressive and energetic species, destroying populations of their preys, but also eliminating species, which cannot successfully compete for the food and thus providing sufficient population reproduction. Preserving such a dynamically balanced state of food chains is the only way for the entire living world to survive as a whole. (If people could learn from this Nature wisdom, it would be for sure a better world for humans.)

So, we can summarize that dynamic balancing and rebalancing of a food chain and its parts is based on at least two principles:

(a) The food chain is continuous, and, if broken, tends to restore its continuity;
(b) Biochemical mechanisms, bio-mechanical constraints and different physical "denominators" (meaning parameters, characterizing the interaction of different species in the food chain) were evolutionarily developed as a statistical summation of adaptations of individual organisms in common nutritional environments in such a way that they allow organisms to adapt to a very wide spectrum of different environmental conditions, far exceeding constraints imposed by particular physiological mechanisms. This entire process evolves within the context defined by natural selection guided by optimization of distribution of resources, available within a common habitat, across the food chain, and this process is the primary cause defining phenomenon of allometric scaling. Optimization of the distribution of shared resources is performed in such a way that each species within the food chain receives the amount of resources for reproduction, sufficient to support the continuation of a food chain, while at the same time not overexploit the common resources to jeopardize the reproduction of other species within the food chain.

Figure 6 represents the above considerations in a schematic form.

### 7.1. *Relationship between the obtained and previous results*

Here, we return to the graph of metabolic rate versus organismal mass from Ref. 26, which we earlier mentioned. The main specific feature of this graph is the high dispersion of presented data in all size ranges. At first glance, it can be viewed as only an obstacle to obtaining accurate regression lines. In fact, such great dispersion is an *inherent* property of the metabolism of unicellular organisms, as it follows from the obtained results, one of which is the necessarily high adaptability of unicellular organisms to different environmental conditions. In particular, it manifests itself as the wide range of nutrient influx per unit surface in and between different organisms, while not superseding the overall trend in the increase of this nutrient influx when organisms become bigger. For instance, as Fig. 5 shows, the mode of motion is an important factor, contributing to such a dispersion, besides other factors, stimulating the high adaptability of unicellular organisms, including the wide range of their metabolic characteristics, and, accordingly, high variability of allometric scaling scenarios, which the aforementioned graph from Ref. 26 confirms.

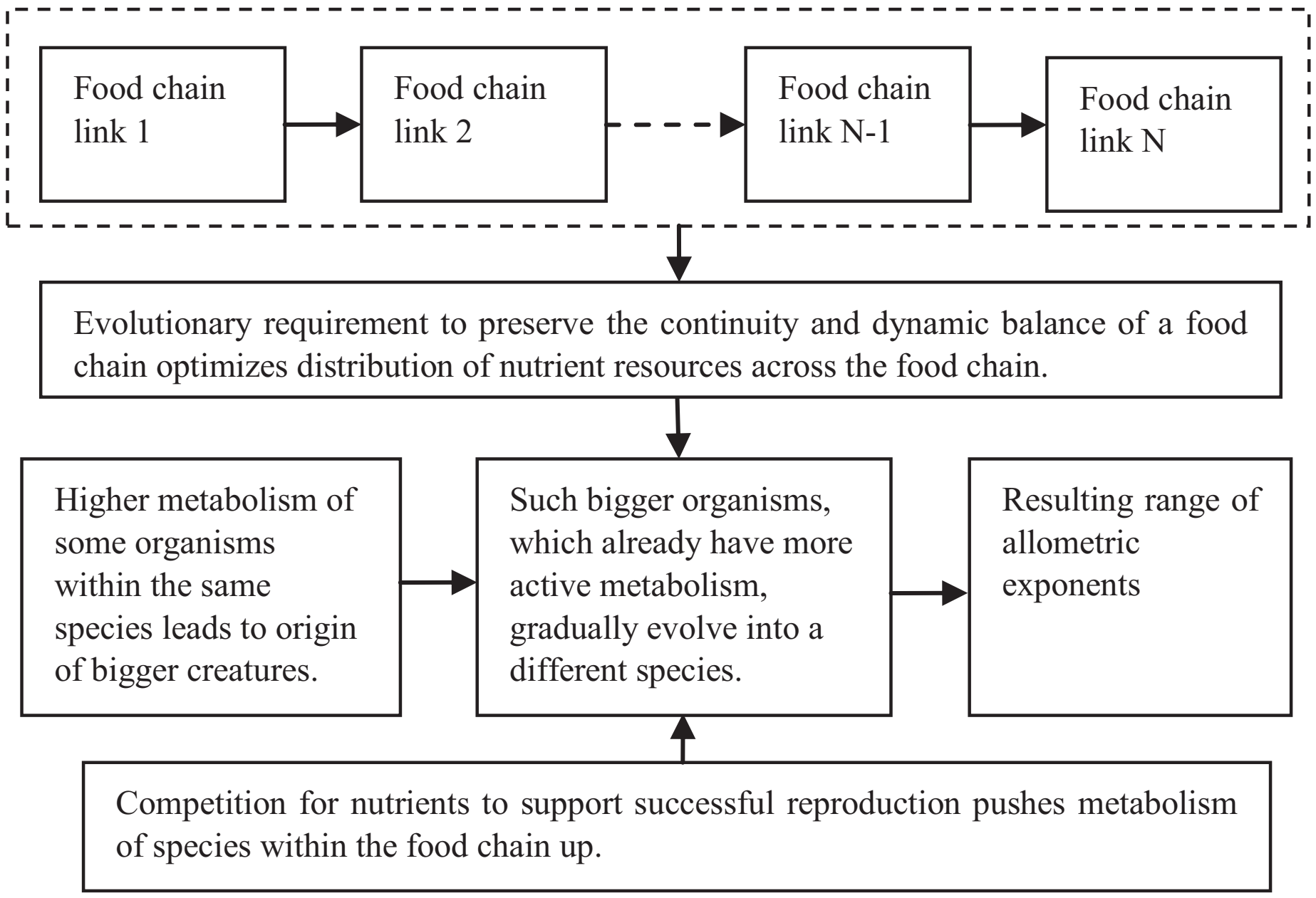

Fig. 6. Evolutionary factors defining metabolic properties of living organisms and interspecific allometric scaling. The requirement of continuity and preserving a dynamic balance of a food chain eliminates too aggressive species, destroying populations of their prey, but also species incapable of supporting sufficient quantities of their populations.

Thus, in this regard, our results correspond to previous studies, which showed a wide range of allometric exponents for unicellular organisms. As it turned out, such high variability of results in the earlier studies is not necessarily the drawback of used methods and approaches, but rather an inherent property of the metabolism of unicellular organisms.

*Validity of assumption about 3D increase of unicellular organisms*

Above, we used the assumption about the 3D increase of species along the food chain. The following considerations support this assumption. Unicellular organisms demonstrate a clear 3D increase when size increases. Indeed, *B. subtilis*, *E. coli*, *S. pombe* are all rod-like, with semispherical ends and close proportions between the lengths and diameters. The same can be said about other geometrical forms, like spheres, or spirals. Given the high level of metabolic adaptability of unicellular organisms, which we found (see Fig. 4 and Table 1), the specifics of their metabolism, like distribution of transportation and nontransportation costs is not so important, since these are rather evolutionary pressures, which trim the metabolic properties of unicellular organisms, but not their particular intrinsic mechanisms. So, despite the specifics of nutrient consumption, the metabolic properties of all organisms will be lined up by the evolutionary development of the entire food chain, which preserves its continuity and a proper dynamic balance.

*The concept of a balanced food chain as a factor shaping allometric scaling, and its verification*

It turned out that finding a solution to the problem of interspecific allometric scaling required the introduction of ideas and concepts, which went beyond particular physiological mechanisms, and, in fact, rose to high-level generalizations related to systemic evolution, that is evolution of a food chain within a given habitat. We formulated a concept of an evolutionary development, constrained by a requirement of a balanced state of an entire food chain, and presented proof that this fundamental evolutionary principle, indeed, shapes different characteristics of living organisms, including metabolic ones.

This principle, by and large, stems from the need to choose between stable and unstable developmental scenarios. The alternative to balance is an imbalance, which would mean frequent, chaotic and sporadic appearance of new species and the disappearance of existing ones in the food chain. However, this is not what we observe in nature, which demonstrates rather stability of its development during long geological periods, until the environmental conditions drastically change or catastrophes happen.

### 7.2. *Future studies*

A more detailed excurse can be done in the discovered facts of regular increase of metabolic power, such as through a certain speed increase per a certain increase in mass, or a certain increase in nutrient influx per unit surface in larger unicellular organisms. We may ask, why namely *such* a certain value of allometric exponent corresponds to a balance, but not another? This is an interesting question on its own. For animals, we did ballpark calculations, what had to be the running distance to catch the prey, using the speed advantage found from the studies, when starting at a certain distance from the prey (meaning a typical ambush distance). Results are in synch with distances, estimated based on several videos, showing chasing animals. The next step would be estimation, of how much energy, and in which form, is stored in an animal to cover a certain distance at a maximum speed. One could do similar estimations for unicellular organisms, using the obtained value of the nutrient influx's increase per unit surface with the size increase, and finding out, what kind of advantages such an increase could provide, and which consequences it entails with regard to successful reproduction.

## 8. Conclusion

This study of interspecific allometric scaling of unicellular organisms, in the author's view, advanced understanding of the phenomenon. The nature of core things in the world we live in is not very complicated, neither it is too simple, as it usually turns out when this core nature is finally discovered. This is what we observed in our study, removing layer after layer before we saw a bare phenomenon without prejudice. The presented causes of the phenomenon of interspecific allometric scaling are relatively simple and noncontradictive. At the same time, one needs to accept things, such as the evolution of the whole food chain within a given habitat, which were not supposed to be there. Such findings make the obtained results to be viewed with some disbelief. The obvious things were that unicellular organisms acquire nutrients *through the surface* (the first important consideration), and from the *common* nutritional environment (the second key consideration, which exposes and raises the role of the environment as a communication media for cells, through which they interact and receive feedbacks). It follows from these considerations that the metabolic properties of cells we studied should be limited by the nutritional content of the environment they reside in, which may be almost obvious, but not easily acceptable inference, given the history of the problem, whose solution was searched for over a wide range of different hypotheses, amongst which the intrinsic factors were of prominence.

The major obstacle in accepting the presented results is the recognition of a concept that the interspecific allometric scaling relates to a balanced existence of the food chain, which accordingly results in constrained competition for nutrient acquisition — no species can dominate absolutely, but always experience outside pressures and limitations in nutrient supply. The last suggestion was not supposed to be in the scope of the problem as well.

The study did not present solid statistical evidence for all considered species. For that, we would need experimental growth dependencies (biomass increase from the growth time), which are not available. On the other hand, the most meaningful range of sizes is between *S. pombe* and *Amoeba*. For the former, we provided statistically meaningful data. For *Amoeba*, we presented one point, which is a good representative of the average value for six available measurements, which all produced close values of nutrient influxes per unit surface and per unit volume. Six measurements is not a great number, of course, but the closeness of calculated nutrient influxes adds more credibility to our estimations. The main value of the study is in the proposed concepts, while experimental data provide convincing support.

The most intriguing result of this study, and of the previous study of metabolic properties of multicellular organisms, is the following development of the general evolution theory. It was found that natural selection has one more important determinant, at the next organizational level, which is a dynamic balance of a food chain in a given habitat, performed through the optimization of sharing of nutritional resources between the species of a food chain. This optimal balance is characterized by well defined and physiologically meaningful quantitative characteristics, such as metabolic allometric exponents, to which many other parameters and developmental processes can be related, at different organizational levels, that is at the levels of individual organisms, species, and of entire food chains.

We think that this work opens new venues for the studies of metabolic allometric scaling, and maybe allometric scaling in general. Together, the suggested methods and concepts provide a framework to make future studies well-designed, planned and conceptually organized. Even more important is that this framework allows understanding and explaining observed effects, and predicting the new ones.

On a side note, the proposed concept of a dynamic balance of a food chain unexpectedly establishes new connections of the subject of allometric scaling with many other disciplines and areas, transforming it from a relatively isolated problem to an important component of numerous evolutionary and biological processes and phenomena. That could be an interesting development too.

## Acknowledgments

The author thanks Dr. A. Y. Shestopaloff for fruitful discussions and help with editing, and Dr. K. Y. Shestopaloff for the help with data processing. Invitation from Editors of *Biophysical Reviews and Letters* to submit papers is very much appreciated, as well as an efficient editorial process. Reviewers' comments helped to improve the paper's structure and its readability.

## ORCID

Yuri K. Shestopaloff https://orcid.org/0000-0003-4251-0618